\documentclass[superscriptaddress,showpacs,amssymb,10pt,reprint,aps,prd,longbibliography,nofootinbib,floatfix]{revtex4-1}

\usepackage{graphicx,epsfig,amssymb}
\usepackage{amsmath,amsfonts, times}
\usepackage{bm}

\usepackage{epstopdf}
\usepackage[linktocpage,colorlinks]{hyperref}
\usepackage[caption=false]{subfig}
\usepackage[usenames]{color}
\usepackage{natbib}
\usepackage{soul}
\usepackage[utf8x]{inputenc}
\usepackage[usenames]{color}

\begin{document}
 \newcommand{\bq}{\begin{equation}}
 \newcommand{\eq}{\end{equation}}
 \newcommand{\bqn}{\begin{eqnarray}}
 \newcommand{\eqn}{\end{eqnarray}}
 \newcommand{\nb}{\nonumber}
 \newcommand{\lb}{\label}
\newcommand{\PRL}{Phys. Rev. Lett.}
\newcommand{\PL}{Phys. Lett.}
\newcommand{\PR}{Phys. Rev.}
\newcommand{\PRD}{Phys. Rev. D.}
\newcommand{\CQG}{Class. Quantum Grav.}
\newcommand{\JCAP}{J. Cosmol. Astropart. Phys.}
\newcommand{\JHEP}{J. High. Energy. Phys.}
\title{\large {Theoretical and observational constraints on regularized 4$D$ Einstein-Gauss-Bonnet gravity}}

	\author{Jia-Xi Feng}
	\email{fengjiaxigw@gmail.com}
	\affiliation{Department of Physics, Nanchang University, Nanchang, 330031, China}
	\affiliation{Center for Relativistic Astrophysics and High Energy Physics, Nanchang University, Nanchang,
330031, China}
	\author{Bao-Min Gu}
	\email{gubm@ncu.edu.cn}
	\affiliation{Department of Physics, Nanchang University, Nanchang, 330031, China}
	\affiliation{Center for Relativistic Astrophysics and High Energy Physics, Nanchang University, Nanchang,
330031, China}
		\author{Fu-Wen Shu}
	\email{shufuwen@ncu.edu.cn; Corresponding author}
	\affiliation{Department of Physics, Nanchang University, Nanchang, 330031, China}
	\affiliation{Center for Relativistic Astrophysics and High Energy Physics, Nanchang University, Nanchang,
330031, China}
\affiliation{Center for Gravitation and Cosmology, Yangzhou University, Yangzhou, China}

\begin{abstract}
Regularized Einstein-Gauss-Bonnet (EGB) theory of gravity in four dimensions is a new attempt to include nontrivial contributions of Gauss-Bonnet term. In this paper, we make a detailed analysis on possible constraints of the model parameters of the theory from recent cosmological observations, and some theoretical constraints as well.  Our results show that the theory with vanishing bare cosmological constant, $\Lambda_0$, is ruled out by the current observational value of $w_{de}$, and the observations of GW170817 and GRB 170817A as well.  For nonvanishing bare cosmological constant, instead, our results show that the current observation of the speed of GWs measured by GW170817 and GRB 170817A would place a constraints on $\tilde{\alpha}$, a dimensionless parameter of the theory, as $-7.78\times10^{-16}\le \tilde{\alpha} \leq 3.33\times10^{-15}$.
	\end{abstract}
	
	\date{\today}
	
	\maketitle

\section{Introduction}
The first detection of gravitational waves (GWs) by LIGO/Virgo \cite{GW150914} begins to have a profound impact on our understanding of the nature. They provide new powerful ways to explore physics of the Universe.
GW170817 \cite{TheLIGOScientific:2017qsa}, the first detected GW event with electromagnetic counterparts, extensively enriched the ways. From then on, a new era of multi-messenger GW astronomy has began. Fermi Gamma-Ray Burst Monitor \cite{Goldstein:2017mmi} and the International Gamma-Ray Astrophysics Laboratory \cite{Savchenko:2017ffs} observed a gamma ray burst GRB 170817A  after $1.74\pm 0.05$s,
on which a range of constraint on the speed of GWs can be  obtained. Particularly, with the assumption that the GW signal was emitted at most 10s before the GRB signal,  one can obtain a bound on the velocity of the GWs, namely, $-3\times 10^{-15}\le \frac{c_{\rm gw}}c-1\le 7\times 10^{-16}$ \cite{Monitor:2017mdv}.

The observations of GW events \cite{GW150914,Abbott:2016nmj,Abbott:2017vtc,Abbott:2017gyy,Abbott:2017oio,TheLIGOScientific:2017qsa} in recent years, of course, support the validity of Einstein's theory enough. However, whether alternative theories of gravity which can do equally well as  Einstein's theory can be constructed or not? This point has attracted a large number of researchers to study, such as  scalar-tensor theories \cite{Horndeski:1974wa,Fujii:2003pa,Chow:2009fm,Tsujikawa:2010zza,Chen:2010va,Clifton:2011jh,Gleyzes:2014qga,Crisostomi:2016czh,Sakstein:2017xjx,Ezquiaga:2017ekz,Green:2017qcv,Casalino:2018wnc}, vector-tensor theories \cite{Baker:2017hug}, and so on. With more and more GW events to be detected in the future, it is expected that constraints on the speed of GWs will be more and more stringent. This makes it an effective tool to test the alternative theories of gravity \cite{Mirshekari:2011yq,Jimenez:2015bwa,Chesler:2017khz,Baker:2017hug,Creminelli:2017sry,Sakstein:2017xjx,Ezquiaga:2017ekz,Green:2017qcv,Nishizawa:2017nef,Arai:2017hxj,Battye:2018ssx}.

Hence, we are paying attention to modified theories of gravity. One of the most elegant modifications is the Einstein-Gauss-Bonnet gravity. It is generally discussed that the extension of higher derivatives by adding the polynomial invariants of the Riemann tensor to the Einstein-Hilbert action is admitted in Einstein's gravity. The field equations which involve four derivatives may lead to renormalizability, while the theory contains an inevitable ghostlike massive graviton \cite{Stelle:1976gc}.
It was found that the Gauss-Bonnet(GB) term, which is a quadratic combination of the Riemann curvature tensor, keeps the equations at second-order in the metric and hence is free of the ghost \cite {Lovelock:1971yv,Zumino:1985dp}. In four or lower dimensions, however, these specific combinations of tensor polynomials either vanish or become total derivative. The trivialness in four dimensions excludes it as a more realistic model.

In \cite{Glavan:2019inb}, a new theory called 4 dimensional Einstein-Gauss-Bonnet ($4D$ EGB) gravity was proposed. It considers a $D\to4$ limit of  the D-dimensional Gauss-Bonnet gravity  by rescaling the GB dimensional coupling constant $\alpha \to \hat{\alpha}/(D-4)$. The idea is to introduce the divergent coefficient to cancel the vanishing contribution of $\mathcal{G}$ in four dimensions, in a manner that is conceptually similar to the dimensional regularization procedure used in quantum field theories. The goal of this is to produce a nontrivial gravity theory in four dimensions that includes a non-vanishing contribution from the Gauss-Bonnet term. A large number of relevant works has been done in the past few months \cite{Nojiri:2020tph,Konoplya:2020bxa,Guo:2020zmf,Fernandes:2020rpa,Casalino:2020kbt,Konoplya:2020qqh,Hegde:2020xlv,Ghosh:2020vpc,Doneva:2020ped,Zhang:2020qew,Konoplya:2020ibi,Singh:2020xju,Ghosh:2020syx,Konoplya:2020juj,Kumar:2020uyz,Zhang:2020qam,HosseiniMansoori:2020yfj,Wei:2020poh,Singh:2020nwo,Churilova:2020aca,Islam:2020xmy, Mishra:2020gce,Kumar:2020xvu,Liu:2020vkh,EGB29,Konoplya:2020cbv,Heydari-Fard:2020sib,Jin:2020emq,Zhang:2020sjh,EslamPanah:2020hoj,Aragon:2020qdc,Yang:2020czk,Lin:2020kqe,Yang:2020jno,Narain:2020qhh,Narain:2020tsw,Ge:2020tid,Banerjee:2020dad}.
However, the resulting theory has been questioned a lot.  The theory is found to be not well defined in the limit $D\rightarrow4$  \cite{Ai:2020peo,Gurses:2020ofy,Lu:2020iav,Kobayashi:2020wqy, Hennigar:2020lsl,Fernandes:2020nbq,Mahapatra:2020rds}. Moreover, the vacua of the model are unstable or ill-defined too \cite{shu}.  To overcome this, several regularization schemes have been proposed \cite{Lu:2020iav,Kobayashi:2020wqy, Hennigar:2020lsl,Fernandes:2020nbq,Aoki:2020iwm}.  This generally leads to a scalar-tensor gravity, being a subclass of Hordenski theories \cite{VanAcoleyen:2011mj}.

In this work, we will perform a detailed analysis of cosmological perturbations of the regularized model around the FRW universe. The speed of tensor modes can be read off from these perturbative equations. Then we look for possible constraints on the coupling constant $\hat{\alpha}$ of the regularized model through latest observational constraints on the speed of GWs  from GW170817 \cite{TheLIGOScientific:2017qsa} and GRB 170817A \cite{Monitor:2017mdv}. Our results show that, for the theory with the bare cosmological constant $\Lambda_0=0$, there are two contradictions: one is the theoretical requirements of the model are contradicted with the current observational value of $w_{de}$. The other is the constraints imposed from GW170817 and GRB 170817A disagree with the current cosmological constraint on the ratio of energy densities between dark energy and matter, $\frac{\Omega_{de}}{\Omega_m}$.  Therefore, the theory $\Lambda_0=0$ is ruled out.  The case with $\Lambda_0\neq0$, however, receives a constraint from the speed of GWs measured by GW170817 and GRB 170817A, explicitly, $-7.78\times10^{-16}\le \tilde{\alpha} \leq 3.33\times10^{-15}$. To see whether the scalar perturbations will give more stringent constrains or not, we discuss scalar perturbations as well\footnote{It is worth noting that  stability of black holes also places constraints on the coupling constants as discussed in \cite{Konoplya:2020der}, where the threshold value of (in)stability $\alpha_{inst}$ are obtained from different orders of the Lovelock theory.
}.

The rest of the paper is organized as follows. In section II, we briefly review the regularized Einstein-Gauss-Bonnet theory in $4$ dimensions with cosmological constant. After applying it to the FRW universe, a set of dynamical equations are obtained, followed by a set of cosmological solutions. In section III, we perform linear perturbation analysis around FRW background. The quadratic action and the velocity of gravitation waves are obtained. In section IV, we apply the observational constraints from GW170817 and GRB 170817A to restrict the coupling constant $\hat{\alpha}$ of the model. In section V, the constrains from the scalar perturbations which may be more stringent are discussed as well. A brief concluding remark is drawn in the last section.

\section{Regularized
Einstein-Gauss-Bonnet Theory in Four Dimensions}\label{CD qim}
The action of Einstein-Gauss-Bonnet theory in $D$ dimensions with cosmological constant is
\bqn
\lb{action}
 S&=& \int_{\mathcal{M}} \mathrm{d}^D x \sqrt{-g} \left(\ R-2\Lambda_0 + \alpha \mathcal{G} \right) + S_m,
\eqn
where $\alpha$ is a coupling constant,  $S_m$ is the action associated with matter field, and the Gauss-Bonnet term is
\bqn
\lb{gravity field eq}
 \mathcal{G}&=& R^2-4R_{\mu\nu}R^{\mu\nu}+R_{\mu\nu\alpha\beta}R^{\mu\nu\alpha\beta}~.
\eqn
The idea of \cite{Glavan:2019inb} is to construct a nontrivial theory by considering a replacement $\alpha \to \frac{\hat{\alpha}}{(D-4)}$. This, however, turns out to be questionable in many aspects. In particular, it was found the theory defined in this way has no well-defined limit \cite{Ai:2020peo,Gurses:2020ofy,Lu:2020iav,Kobayashi:2020wqy, Hennigar:2020lsl,Fernandes:2020nbq,shu}. The way to fix this pathology is to perform a regularization. There are several regularization schemes in the literatures, such as the Kaluza–Klein-reduction procedure \cite{Lu:2020iav,Kobayashi:2020wqy}, the conformal subtraction procedure \cite{Hennigar:2020lsl,Fernandes:2020nbq}, and ADM decomposition analysis \cite{,Aoki:2020iwm}. The first two approaches give rise to the same regularized action, which is of the following form\footnote{This action belongs to a subclass of the Horndeski gravity  \cite{Horndeski:1974wa,Kobayashi:2019hrl}  with  $G_2=8 \hat{\alpha} X^2-2\Lambda_0$, $G_3=8 \hat{\alpha} X$, $G_4=1+4 \hat{\alpha} X$ and $G_5 = 4 \hat{\alpha} \ln X$ \big(where $X=-\frac{1}{2} \nabla_{\mu} \phi \nabla^{\mu} \phi$\big).}

\bqn
\lb{EGBaction}
 S = \int_{\mathcal{M}} \mathrm{d}^4 x \sqrt{-g} \Big[&R& -2\Lambda_0+ \hat{\alpha} \Big(4G^{\mu \nu}\nabla_\mu \phi \nabla_\nu \phi - \phi \mathcal{G}\nb\\
 && + 4\Box \phi (\nabla \phi)^2 + 2(\nabla \phi)^4\Big) \Big] + S_m,\nb\\
\eqn
where $\phi$ is a scalar field inherent from $D$ dimensions. It is introduced by Kaluza–Klein reduction of the metric
\cite{Lu:2020iav,Kobayashi:2020wqy}
$$
\mathrm{d}s_D^2=\mathrm{d}s_4^2+e^{2\phi}\mathrm{d}\Omega_{D-4}^2,
$$
or by conformal subtraction \cite{Hennigar:2020lsl,Fernandes:2020nbq} where the subtraction background is defined under a conformal transformation $g_{ab}\rightarrow e^{2\phi}g_{ab}$.

Varying with respect to the metric, we can get the field equations
 \bqn
\lb{gravity field eq}
 G_{\mu \nu}+\Lambda_0 g_{\mu \nu} =  \hat{\alpha} \hat{\mathcal{H}}_{\mu \nu} +T_{\mu \nu}^{(m)},
\eqn
where $T_{\mu \nu}^{(m)}$ is the energy-momentum tensor of the matter field, considering the matter context of the universe is a perfect fluid, so that the energy-momentum tensor take the form \cite{Carroll:2004st}
\bqn
\lb{cc}
T_{\mu \nu}^{(m)} &=& (\rho_m +p_m)U_{\mu}U_{\nu} +p_mg_{\mu \nu},
\eqn
where  $\rho_m$,  $p_m$  and  $U_{\mu} $ are respectively energy density, pressure and four-velocity of the fluid. And

 \begin{widetext}
\bqn
\lb{H}
\hat{\mathcal{H}}_{\mu\nu} &=& 2R\big(\nabla_\mu \nabla_\nu \phi - \nabla_\mu\phi \nabla_\nu \phi\big) + 2G_{\mu \nu}\Big(\big(\nabla \phi\big)^2-2\Box \phi\Big)+ 4G_{\nu \alpha} \big(\nabla^\alpha \nabla_\mu \phi -\nabla^\alpha \phi \nabla_\mu \phi\big)\nb\\
 &&+ 4G_{\mu \alpha} \big(\nabla^\alpha \nabla_\nu \phi - \nabla^\alpha \phi \nabla_\nu \phi\big) + 4R_{\mu \alpha \nu \beta}\big(\nabla^\beta \nabla^\alpha \phi - \nabla^\alpha \phi \nabla^\beta\phi\big)\nb\\
 &&+ 4\nabla_\alpha\nabla_\nu \phi \big(\nabla^\alpha \phi \nabla_\mu \phi - \nabla^\alpha \nabla_\mu \phi \big)+4 \nabla_\alpha \nabla_\mu \phi \nabla^\alpha\phi \nabla_\nu \phi  - 4\nabla_\mu \phi \nabla_\nu \phi \Big(\big(\nabla \phi\big)^2+ \Box \phi \Big)\nb\\
 &&  +4\Box \phi\nabla_\nu \nabla_\mu \phi- g_{\mu \nu} \Big[ 2R\big(\Box \phi  - (\nabla \phi)^2\big) + 4 G^{\alpha \beta} \big( \nabla_\beta \nabla_\alpha \phi -\nabla_\alpha \phi \nabla_\beta \phi \big) + 2\big(\Box \phi \big)^2 \nb\\
 && - \big( \nabla \phi\big)^4 + 2\nabla_\beta \nabla_\alpha\phi \big(2\nabla^\alpha \phi \nabla^\beta \phi - \nabla^\beta \nabla^\alpha \phi \big)\Big]\nb\\
\eqn
\end{widetext}

By varying with respect to the scalar field, we get
 \bqn
\lb{scalar field}
\frac{1}{8}\mathcal{G} &=&  R^{\mu \nu} \nabla_{\mu} \phi \nabla_{\nu} \phi - G^{\mu \nu}\nabla_\mu \nabla_\nu \phi - \Box \phi (\nabla \phi)^2\nb\\
 &&  +(\nabla_\mu \nabla_\nu \phi)^2- (\Box \phi )^2 - 2\nabla_\mu \phi \nabla_\nu \phi \nabla^\mu \nabla^\nu \phi\nb\\
\eqn
 The trace of the field equations (\ref{gravity field eq}) is found to satisfy
\bqn
\lb{GB3}
R+\frac{\hat{\alpha}}{2} \mathcal{G}-4\Lambda_0 = -T,
\eqn
where $T=-\rho_m+3p_m$.\\\

Assuming that the line-element describing by spatially-flat Friedmann-Robertson-Walker (FRW) metric is
\bq
\mathrm{d}s^2 = -\mathrm{d}t^2+a^2(t)(\mathrm{d}x^2_1+\mathrm{d}x^2_2+\mathrm{d}x^2_3),
\eq
then taking a direct calculation, we show that the equations of motion become
\bqn
\lb{eom1}
3H^2&=&\rho_{GB}+\rho_m+\rho_{\Lambda},\nb \\
2\dot{H}+3H^2&=&-p_{GB}-p_m-p_{\Lambda},
\eqn
where the energy density, pressure of the GB term and the cosmological constant term are defined as
\bqn
\lb{GB1}
\rho_{GB}&\equiv&\Big(3\dot{\phi}^4+12\dot{\phi}^3H+18\dot{\phi}^2H^2+12\dot{\phi}H^3\Big)\hat{\alpha},\nb \\
p_{GB}&\equiv&\Big[\dot{\phi}^4-4\dot{\phi}^2\ddot{\phi}-4\dot{\phi}^2(\dot{H}+H^2)-2\dot{\phi}^2H^2\nb\\
 &&-8\dot{\phi}\ddot{\phi}H-8\dot{\phi}\big(\dot{H}+H^2\big)H-4\ddot{\phi}H^2\Big]\hat{\alpha},\nb\\
 \rho_{\Lambda}&\equiv&\Lambda_0,\nb\\
 p_{\Lambda}&\equiv&-\Lambda_0.
\eqn
And the scalar field equation, which is equivalent to $\nabla^\mu\hat{\mathcal{H}}_{\mu\nu} =0$ ,  reduces to,
\bq
\partial_t [(a\dot{\phi}+\dot{a})^2]=0,
\eq
which can be solved simply by
 \bqn
 \lb{Solution1}
\dot{\phi}&=&-H,
\eqn
or
 \bqn
 \lb{Solution2}
\dot{\phi}&=&-H+\frac{A}{a},
\eqn
where  $H$ is the Hubble parameter, dot denotes derivative with respect to $t$, $A$ is the integration constant. It is obvious that these solutions are similar with \cite{Lu:2020iav,Kobayashi:2020wqy}.

\section{The speed of gravitational waves }\label{HT}
To study the gravitational waves, let us consider the linear tensor perturbations of the FRW metric,
\bqn
\lb{tensor:per}
\mathrm{d}s^2=-\mathrm{d}t^2+a^2 \left(t\right)\left(\delta_{ij}+h_{ij}\right)\mathrm{d}x^i \mathrm{d}x^j,
\eqn
where the tensor $h_{ij}$ satisfies the transverse-traceless condition, $\partial^ih_{ij}=0=\delta^{ij}h_{ij}$.
Then the linear order field equation of $h_{ij}$ can be expressed as
\begin{equation}
\beta_1 \ddot{h}_{ij}+\left(\dot{\beta_1}+3H\beta_1\right)\dot{h}_{ij}
-\beta_2\frac{\vec{\nabla}^2}{a^2}h_{ij}=0.
\end{equation}
The coefficients $\beta_1$ and $\beta_2$ are defined as
\begin{eqnarray}
&&\beta_1\equiv1-2\hat{\alpha}\dot{\phi}^2-4\hat{\alpha}\dot{\phi}H,
\\
&&\beta_2\equiv1+2\hat{\alpha}\dot{\phi}^2-4\hat{\alpha}\ddot{\phi}.
\end{eqnarray}
The corresponding quadratic action is
\begin{equation}
S_h=\int \mathrm{d}t \mathrm{d}^3x a^3\left[\beta_1 \dot{h}_{ij}^2- \frac{\beta_2}{a^2}\left(\vec{\nabla}h_{ij}\right)^2\right].
\end{equation}
To avoid ghost and gradient instability \cite{Kobayashi:2019hrl}, the two coefficients $\beta_1$ and $\beta_2$ should be positive, namely, $\beta_1>0$ and $\beta_2>0$. This imposes constraints on the coupling constant $\hat{\alpha}$, and we will recall these constraints in next section.

It is more convenient to make the Fourier transformation and write the tensor perturbation $h_{ij}(\vec{x},t)$ as
\begin{equation}\label{FT}
h_{ij}(\vec{x},t)=\int \frac {\mathrm{d}^3 k}{(2\pi)^3}\left[h_k^+\left(t\right)A^+_{ij}+h_k^{\times}\left(t\right)A^{\times}_{ij}\right]
\exp\left(i \vec{k}\cdot \vec{x}\right),
\end{equation}
where $k^i A_{ij}^{\sigma}=A_{ii}^{\sigma}=0$, $A_{ij}^{\sigma}A_{ij}^{\hat{\sigma}}=2\delta_{\sigma\hat{\sigma}}$,
and the superscript ``$\sigma$" stands for the ``$+$" or ``$\times$" polarizations. In terms of the Fourier modes,
we have
\begin{equation}
\label{eq2}
  \ddot{h}_k^{\sigma}+\left(3+\alpha_M\right)H\dot{h}_k^{\sigma}+\frac{c^2_{\rm gw} k^2}{a^2}h_k^{\sigma}=0,
\end{equation}
where
\bqn
\lb{alpha_M}
\alpha_M=\frac{\dot{\beta}_1}{H\beta_1},\quad
c_{\rm gw}^2 =\frac{\beta_2}{\beta_1},
\eqn
$c_{\rm gw}$ is actually the propagation speed of the gravitational waves, $\alpha_M$ is a dimensionless parameter which describes the the running of the effective Planck mass.
We see that there are  modifications to the Hubble friction and the gravitational wave speed. Using the expressions of $\beta_1$ and $\beta_2$, we get
\bqn
\lb{CGWT}
\alpha_M &=&\frac{-4\hat{\alpha}\dot{\phi}\ddot{\phi}-4\hat{\alpha}\dot{\phi}\dot{H}-4\hat{\alpha}\ddot{\phi}H}{H\big(1-2\hat{\alpha}\dot{\phi}^2-4\hat{\alpha}\dot{\phi}H\big)},\nb\\
c_{\rm gw}^2 &=&\frac{1+2\hat{\alpha}\dot{\phi}^2-4\hat{\alpha}\ddot{\phi}}{1-2\hat{\alpha}\dot{\phi}^2-4\hat{\alpha}\dot{\phi}H}.
\eqn
Recalling the solution \eqref{Solution1}, we find that both $\alpha_M$ and $c_{\rm gw}$ are functions of $H$ and its derivative, whose form depends on cosmological models as we will see in the next section.

\section{ The effect of the speed of gravitational waves in the
Gauss-Bonnet theory}\label{3}
In this section we would like to consider possible constraints on the Gauss-Bonnet coefficient  $\hat{\alpha}$  from the current cosmological and gravitational waves' observations.

Substituting the solution (\ref{Solution1}) into  $\beta_1$, $\beta_2$  and equations \eqref{CGWT}, we have
\begin{eqnarray}
&&\beta_1=1+2\hat{\alpha}H^2,\label{b1}
\\
&&\beta_2=1+2\hat{\alpha}H^2+4\hat{\alpha}\dot{H},\label{b2}
\end{eqnarray}
and
\bqn
\lb{alpha_M1}
\alpha_{M} &=&\frac{4\hat{\alpha}\dot{H}}{1+2\hat{\alpha}H^2},
\\
\lb{CGWT1}
c_{\rm gw}^2&=&1+\frac{4\hat{\alpha}\dot{H}}{1+2\hat{\alpha}H^2}.
\eqn
Notice that  the propagation speed of the tensor modes (\ref{CGWT1}) has the same form as \cite{Aoki:2020iwm}. If we take their definition and notation, it can be seen that the propagation speed denoted by  $c_{T}^2$ in their work is equal to $c_{\rm gw}^2$ here. The similar is true for $\alpha_{M}$. In addition, it is worth noting that the dispersion relation obtained here lacks $k^4$ term, which agrees with the prediction made in  \cite{Aoki:2020iwm}. The lack of $k^4$ term means that the present model encounters a strong coupling problem as found in \cite{Bonifacio:2020vbk} and it only captures the IR limit of the full theory proposed in \cite{Aoki:2020iwm}.  Since we are focusing on IR behavior of the theory in this work, we leave this pathology for future work.

Now $\rho_{GB}$ and $p_{GB}$ are given by
\bqn
\lb{GB2.1}
\rho_{GB}&=&-3\hat{\alpha}H^4,\nb \\
p_{GB}&=&\hat{\alpha}H^2\big(3H^2+4\dot{H}\big),
\eqn
then the equations of motion (\ref{eom1}) become
\bqn
\lb{eom1.1}
3H^2&=&\rho_m-3\hat{\alpha}H^4+\Lambda_0,\nb\\
2\dot{H}+3H^2&=&-p_m-\hat{\alpha}H^2\big(3H^2+4\dot{H}\big)+\Lambda_0.
\eqn
In the rest part, two cases will be discussed: one is $p_m = 0$ while $\Lambda_0 \neq 0$, the other is $p_m = 0$ and $\Lambda_0 = 0$. However, the latter case is found to have two contradictory points which will be discussed later.

\subsection{ $p_m = 0$ ,  $\Lambda_0 \neq 0$ }\label{3.1}
Now let us focus on the case where $p_m = 0$ and $\Lambda_0 \neq 0$, describing current acceleration of the universe. Constraints on $\hat{\alpha}$  will be performed at great length in what follows.

First, the energy density of matter can be obtained by the equation of motion  (\ref{eom1.1})
\bqn
\lb{GBed}
\rho_{m}&=&-2\dot{H}(t)-4\hat{\alpha}H^2\dot{H}(t).
\eqn
Defining  a dimensionless  parameter $\tilde{\alpha}$
\bqn
\lb{a}
\tilde{\alpha} &\equiv& \hat{\alpha}H_0^2,
\eqn
then the ratio of the current value of the energy densities between dark energy and matter is given by
\bqn
\lb{rhodedm1}
\frac{\Omega_{de}}{\Omega_m}&=&\frac{\rho_{GB}+\rho_{\Lambda}}{\rho_m}=\frac{3H_0^2-\rho_m}{\rho_m}\nb\\
&=&\frac{3H_0^2+2\dot{H}(t_0)+4\tilde{\alpha}\dot{H}(t_0)}{-2\dot{H}(t_0)-4\tilde{\alpha}\dot{H}(t_0)},
\eqn
where $\rho_{de}=\rho_{GB}+\rho_{\Lambda}$, and $H_0$ is the Hubble constant in present universe. Meanwhile, the current equation of state parameter for dark energy is of the form
\bqn
\lb{wdeeq1}
w_{de}&=&\frac{p_{GB}+p_{\Lambda}}{\rho_{GB}+\rho_{\Lambda}}=\frac{-2\dot{H}(t_0)-3H_0^2}{3H_0^2+2\dot{H}(t_0)+4\tilde{\alpha}\dot{H}(t_0)},
\eqn
where $p_{de}=p_{GB}+p_{\Lambda}$.  The current cosmological observation suggests that the ratio is approximately $\Omega_{de}/\Omega_m=7/3$  \cite{Ade:2015xua} , we hence get
\bqn
\lb{Hdot0}
\dot{H}(t_0)&=&-\frac{9 H_0^2}{20(1+2\tilde{\alpha})}.
\eqn

In what follows we would like to show that four possible constraints on the model parameters $\tilde{\alpha}$ can be imposed, theoretically and observationally.
\begin{itemize}
  \item  Constraints from $\beta_1$ and $\beta_2$

  In section \ref{HT} we have shown that $\beta_1>0$ and $\beta_2>0$ should be satisfied so that the theory is free of ghost and gradient instability. From eqs. \eqref{b1}, \eqref{b2} and \eqref{Hdot0} we have
\begin{eqnarray}
&&\beta_1=1+2\tilde{\alpha},
\\
&&\beta_2=1+2\tilde{\alpha}-\frac{9\tilde{\alpha}}{5(1+2\tilde{\alpha}) }.
\end{eqnarray}
The positivity of $\beta_i$ forces $\tilde{\alpha}$ to be
\bqn
\lb{constraint1}
\tilde{\alpha}>-0.5.
\eqn

  \item  Constraints from  $w_{de}$

Substituting \eqref{Hdot0} into \eqref{wdeeq1}, we then get
\begin{eqnarray}\label{wde}
w_{de}=-\frac{7+20\tilde{\alpha}}{7+14\tilde{\alpha}}.
\end{eqnarray}
This indicates that the observational value of $w_{de}$ will place new constraint on the parameter $\tilde{\alpha}$.
Current cosmological observation shows that $w_{de}$ is bounded by $-1.1<w_{de}<-0.9$  \cite{Ade:2015xua}. Inserting this into \eqref{wde} one has
\bqn
\lb{constraint2}
-0.0945946 < \tilde{\alpha} < 0.152174.
\eqn

  \item  Constraints from $\alpha_M$

The current cosmological constraints on $\alpha_M$ is $-0.434 < \alpha_M < 0.945$ (the parametrization I) at $95\%$ confidence level \cite{Noller:2018wyv}. Eqs. \eqref{alpha_M1} and \eqref{Hdot0} lead to
 \begin{eqnarray}
\alpha_M=-\frac{9\tilde{\alpha}}{5(1+2\tilde{\alpha})^2},
\end{eqnarray}
which implies
\bqn
\lb{constraint3}
\tilde{\alpha} &<&-1.28104,
\eqn
or
\bqn
 \lb{constraint4}
 \tilde{\alpha} &>&-0.195155.
\eqn
However, \eqref{constraint1} shows that the constraint (\ref{constraint3}) should be abandoned.

  \item  Constraints from  GW170817 and GRB 170817A

Thanks to the first detection of an electromagnetic counterpart (GRB 170817A) to the gravitational wave signal (GW170817), we have a new powerful way in testing theories of gravity.  It is well known this event gave rise to a new stringent bound on the speed of GWs has been suggested by using the GW170817 and the GRB 170817A \cite{Monitor:2017mdv}
\bqn
\lb{speed:condition1}
-3\times10^{-15}\le c_{\rm gw}-1\leq 7 \times 10^{-16}.
\eqn
On the other hand, from eqs. \eqref{CGWT1} and \eqref{Hdot0} we have
 \begin{eqnarray}
c^2_{\rm gw}=1-\frac{9\tilde{\alpha}}{5(1+2\tilde{\alpha})^2}.
\end{eqnarray}
This places more constraints on the parameter $\tilde{\alpha}$
\bqn
\lb{constraint5}
-7.78\times10^{-16} \le  \tilde{\alpha} &\leq& 3.33\times10^{-15},
\eqn
or
\bqn
\lb{constraint7}
\tilde{\alpha} &\le & -3.21\times10^{14},
\eqn
or
\bqn
\lb{constraint8}
\tilde{\alpha} &\geq &7.50\times10^{14}.
\eqn
However, the bound (\ref{constraint7}) should be abandoned due to $\beta_1,\beta_2>0$ as shown in \eqref{constraint1}, and the bound (\ref{constraint8}) is also invalid due to (\ref{constraint2}).
\end{itemize}

In summary, combining all these constraints, the latest observations of the speed of GWs from GW170817 and GRB 170817A impose the most stringent one, which is
\bqn
\lb{speed:constraint1}
-7.78\times10^{-16}\le \tilde{\alpha} &\leq& 3.33\times10^{-15}.
\eqn
Note that the expression of $\Lambda_0$ can be obtained from Eqs. (\ref{eom1.1}) and \eqref{Hdot0} as follow
\bqn
\lb{Lambda}
\Lambda_0&=&\Big(3H_0^2+4\dot{H}(t_0)\Big)\tilde{\alpha}+2\dot{H}(t_0)+3H_0^2,\nb\\
&=&\frac{3}{10}\Big(10\tilde{\alpha}+7\Big)H_0^2,
\eqn
which shows that these two parameters are not independent.

\begin{figure}[htb]
\centering
\includegraphics[scale=0.25]{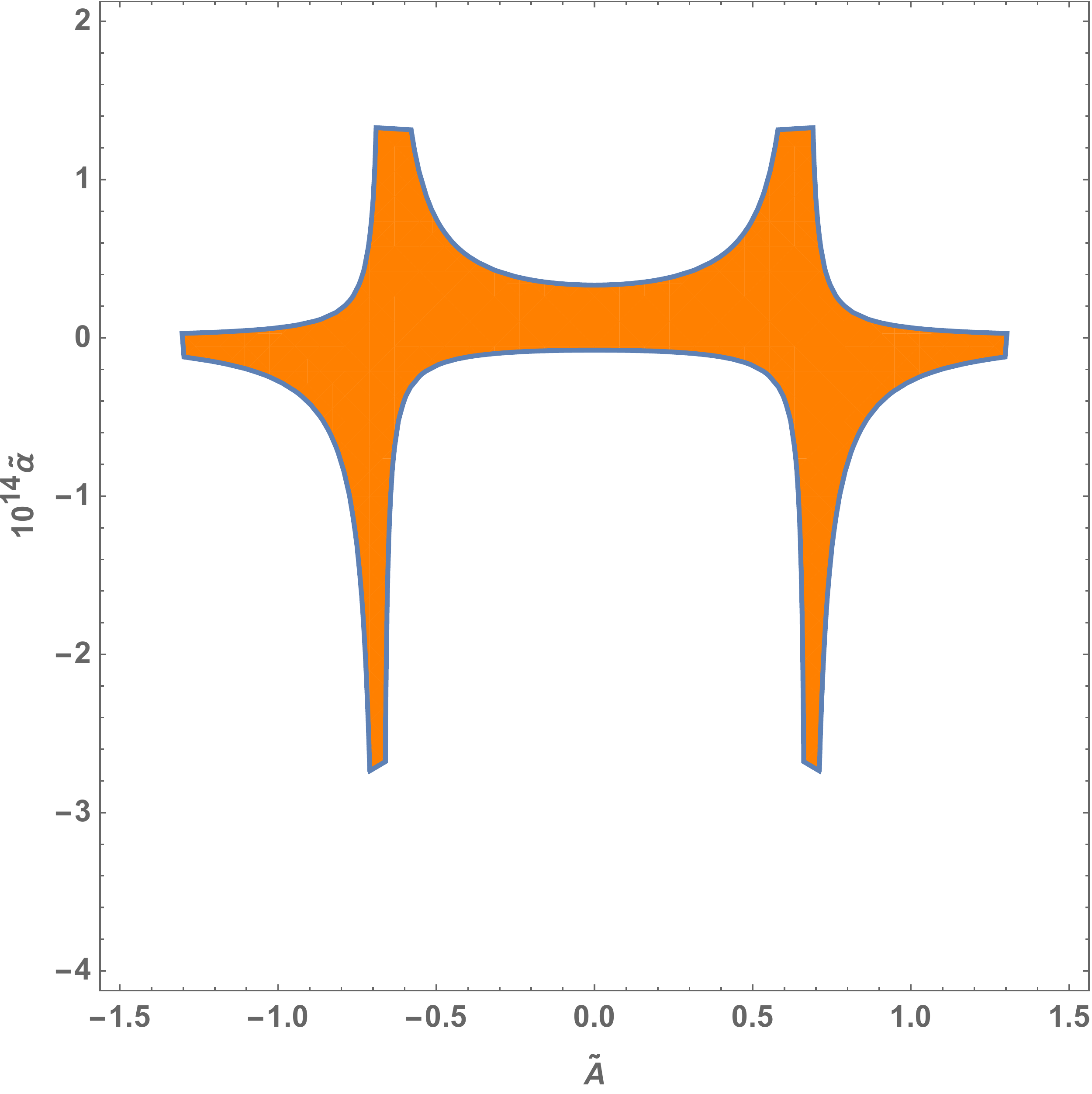}
\caption{Constraints on $\tilde{\alpha}$ and $\tilde{A}$ from theoretical and observational bounds. The orange area is allowed by the constraints mentioned in the main text.} \label{fig2}
\end{figure}

One may expect that including the integration constant $A$ in \eqref{Solution2} may give more stringent constraint. Indeed, for some values of $A$, $\tilde{\alpha}$ receives more stringent constraints. To see this explicitly, let us follow what we did on above.
By defining a dimensionless variable $\tilde{A}\equiv\frac{A}{H_0}$,
$\dot{H}(t_0)$ can be obtained from the current value of $\frac{\Omega_{de}}{\Omega_m}$
\bqn
\lb{Hdot2}
\dot{H}(t_0)&=&-\frac{9+40\tilde{\alpha}\tilde{A}^4}{20\big(1+2\tilde{\alpha}\big)}H_0^2,
\eqn
which recovers the results \eqref{Hdot0} where $A=0$. We then obtain that
\bqn
\label{beta12}
\beta_1&=&1-2\tilde{\alpha}\left(1-\tilde{A}\right)^2+4\tilde{\alpha }\left(1-\tilde{A}\right),\nb\\
\beta_2
&=&\frac{5+\tilde{\alpha }\left(11+10 \tilde{A}^2\right)+ 20 \tilde{\alpha }^2 \left(1+\tilde{A}^2-2 \tilde{A}^4\right)}{5+10 \tilde{\alpha }},\nb\\
\eqn
and
\bqn
\label{wde2}
w_{de}&=&-\frac{21+60\tilde{\alpha}-40\tilde{\alpha } \tilde{A}^4}{21\big(1+2\tilde{\alpha}\big)},\nb\\
\label{am}
\alpha_M &=&-\frac{9 \tilde{\alpha }+40 \tilde{\alpha }^2 \tilde{A}^4-20 \tilde{\alpha }\big(1+2 \tilde{\alpha }\big) \tilde{A}^2}{5 \big(1+2 \tilde{\alpha }\big) \big(1+2 \tilde{\alpha }-2 \tilde{\alpha } \tilde{A}^2\big)},\nb\\
\label{CGWT2}
c_{\rm gw}^2 &=&1-\frac{9 \tilde{\alpha }+40 \tilde{\alpha }^2 \tilde{A}^4-20 \tilde{\alpha }\big(1+2 \tilde{\alpha }\big) \tilde{A}^2}{5 \big(1+2 \tilde{\alpha }\big) \big(1+2 \tilde{\alpha }-2 \tilde{\alpha } \tilde{A}^2\big)}.
\eqn
Using the constraints from theoretical and observational bounds mentioned on above, we can make a plot (Fig. \ref{fig2}). From this plot we find that for $|\tilde{A}|\lesssim 1.35$ we have constraints $|\tilde{\alpha}|<3\times10^{-14}$. In addition, the case with $|\tilde{A}|\gtrsim 1.35$ is excluded by observational data as shown in Fig. \ref{fig2}. Only the value of orange region in Fig. \ref{fig2} is allowed.

\subsection{$p_m = 0$ , $\Lambda_0 = 0$ }\label{3.2}

In this subsection, let us turn to consider the case where  the theory has vanishing bare cosmological constant, namely, $p_m = 0$ and $\Lambda_0 = 0$.

From \eqref{eom1.1}, it is straightforward to show that $\dot{H}(t)$ and $\rho_m$ are, respectively, given by
\bqn
\lb{Hdot1}
\dot{H}(t)&=&-\frac{3(1+\hat{\alpha}H^2 )H^2}{2(1+2\hat{\alpha}H^2 )},\\
\rho_{m}&=&3H^2(1+\hat{\alpha}H^2 ).
\eqn

Just like what we did in $\Lambda_0\neq 0$ case, four possible bounds on the model parameter $\tilde{\alpha}$ can be obtained (where, again, we have introduced a dimensionless  parameter $\tilde{\alpha}=\hat{\alpha}H_0^2$).
\begin{itemize}
 \item  Constraints from $\beta_1$ and $\beta_2$

In the present case, $\beta_1$ and $\beta_2$ become
\begin{eqnarray}
&&\beta_1=1+2\tilde{\alpha},
\\
&&\beta_2=1+2\tilde{\alpha}-6\tilde{\alpha}\cdot\frac{1+\tilde{\alpha}}{1+2\tilde{\alpha}}.
\end{eqnarray}
The requirement that $\beta_i>0$ thus place a constraint on $\tilde{\alpha}$ as
\bqn
\lb{constraint1.1}
-0.5<\tilde{\alpha}<0.366025.
\eqn
  \item  Constraints from  $w_{de}$

 The expression for $w_{de}$ now becomes
\begin{eqnarray}
w_{de}=\frac1{\tilde{\alpha}}\left(1-\frac{1+\tilde{\alpha}}{1+2\tilde{\alpha}}\right).
\end{eqnarray}
The current bound on $w_{de}$ is $-1.1<w_{de}<-0.9$ \cite{Ade:2015xua} implies that
\bqn
\lb{constraint2.1}
-1.0556 < \tilde{\alpha} < -0.954545.
\eqn
Clearly, this result contradicts with (\ref{constraint1.1}), a theoretical requirement to guarantee the theory is free from ghost and instabilities. This strongly suggests that the model with vanishing bare cosmological constant is ruled out from current cosmological observations. In what follows, we will show another evidence to support this statement.

\item  Constraints from $\alpha_M$

From \eqref{CGWT1} and \eqref{Hdot1} one gets
 \begin{eqnarray}
\alpha_M=-\frac{6\tilde{\alpha}(1+\tilde{\alpha})}{(1+2\tilde{\alpha})^2}.
\end{eqnarray}
Again we use the current cosmological constraints of $\alpha_M$, $-0.434 < \alpha_M < 0.945$ \cite{Noller:2018wyv}, then we get
\bqn
\lb{constraint3.1}
-1.09311<\tilde{\alpha}<-0.89163,
\eqn
or
\bqn
 \lb{constraint4.1}
 -0.10837<\tilde{\alpha}<0.0931124.
\eqn
 Clearly the bound (\ref{constraint3.1}) should be abandoned due to $\beta_1,\beta_2>0$ .

\item Constraints from  the speed of GWs

Using the bound on the speed of GWs \cite{Monitor:2017mdv},
\bqn
\lb{speed:condition2}
-3\times10^{-15}\le c_{\rm gw}-1\leq 7 \times 10^{-16},
\eqn
and the expression of $c^2_{\rm gw}$ of this case
 \begin{eqnarray}
c^2_{\rm gw}=1-\frac{6\tilde{\alpha}(1+\tilde{\alpha})}{(1+2\tilde{\alpha})^2},
\end{eqnarray}
we find the following bounds
\bqn
\lb{constraint5.1}
-1\times10^{-15}&\le&\tilde{\alpha}+1\leq 2\times10^{-16},
\eqn
or
\bqn
\lb{speed:constraint2.1}
-2.33\times10^{-16}&\le&\tilde{\alpha}\leq 1.0\times10^{-15}.
\eqn
It is obvious that the bound (\ref{constraint5.1}) is not allowed because of \eqref{constraint1.1}.

\end{itemize}

In summary, if we put the inconsistency obtained from the constraint of  $w_{de}$ aside,  we naively have a stringent bound \eqref{speed:constraint2.1}. However, we should be very careful here. If we take the current cosmological observations into consideration, we find there is an inconsistency in this case. Particularly,
the current cosmological observations put a severe constraint on the ratio of energy densities between dark energy and matter (i.e. $\frac{\Omega_{de}}{\Omega_m}\sim \frac73$). Direct computation shows that the present case leads to the following ratio
\bqn
\lb{rhodedm1}
\frac{\Omega_{de}}{\Omega_m}&=&\frac{\rho_{GB}+0}{\rho_m}=\frac{3H_0^2-\rho_m}{\rho_m}\nb\\
&=&-\frac{\tilde{\alpha}}{1+\tilde{\alpha}},
\eqn
which is much much less than $7/3$ after combining the result \eqref{speed:constraint2.1}. This provides another evidence, in addition to the one given in  (\ref{constraint1.1}) and (\ref{constraint2.1}), for the inconsistency of the model with vanishing bare cosmological constant.

Including the integration constant $A$
in (14), there is same inconsistency obtained from the constraint of $w_{de}$. The region of the constraints from $\beta_1, \beta_2,$ and $w_{de}$ on $\tilde{\alpha}$ and $\tilde{A}$  have no intersection. Hence, we conclude that the model in question does not admit a cosmological solution with vanishing bare cosmological constant, $\Lambda_0 = 0$.

\section{The scalar perturbations}\label{4}
Now let us consider the scalar perturbations. We choose the unitary gauge, in which the fluctuation of the scalar field vanishes and all of the fluctuations are described by that of the spacetime metric.  The line element is assumed as
\begin{equation}
\mathrm{d}s^2=-(1+2\Phi)\mathrm{d}t^2 + 2\partial_i\xi \mathrm{d}t\mathrm{d}x^i+a^2(1+2\Psi)\delta_{ij}\mathrm{d}x^i\mathrm{d}x^j.
\end{equation}
Varying the action with respect to $\Phi$ and $\xi$ leads to two constraints, corresponding to the energy and momentum constraints,
\begin{eqnarray}
\beta_1 \dot{\Psi} + \gamma_1 \Phi&=&0,
\\
\gamma_1 \frac{\vec{\nabla}^2\xi}{a^2}- \beta_1\frac{\vec{\nabla}^2\Psi}{a^2}
-3\gamma_1 \dot{\Psi}+\gamma_2\Phi&=&0.
\end{eqnarray}
Here we introduce two coefficients,
\begin{eqnarray}
&&\gamma_1=6\hat{\alpha} H^2\dot{\phi}+6\hat{\alpha} H\dot{\phi}^2
+2\hat{\alpha}\dot{\phi}^3-H, \\
&&\gamma_2=3H^2-2\Lambda_0.
\end{eqnarray}
The variation of $\Psi$ then gives a nontrivial field equation
\begin{eqnarray}
-H\beta_2 \vec{\nabla}^2\xi-\beta_1\vec{\nabla}^2\dot{\xi}
+3a^2 \beta_1\ddot{\Psi}-\beta_2\vec{\nabla}^2\Psi+\gamma_3\dot{\Psi}
\nonumber\\
-\beta_1\vec{\nabla}^2\Phi+3a^2\gamma_1 \dot{\Phi}+\gamma_4\Phi=0,
\end{eqnarray}
where $\gamma_3$ and $\gamma_4$ are defined by
\begin{eqnarray}
&&\gamma_3=3a\dot{a}(2\beta_1+\beta_2),
\\
&&\gamma_4=3a^2 (3H\gamma_1+\dot{\gamma}_1).
\end{eqnarray}
Using the two constraints we can eliminate $\Phi$ and $\xi$, and get the equation for $\Psi$,
\begin{equation}
\sigma_1\ddot{\Psi}+(3H\sigma_1+\dot{\sigma_1})\dot{\Psi}
-\frac{\sigma_2}{a^2}\vec{\nabla}^2\Psi=0.
\end{equation}
The corresponding quadratic action is
\begin{equation}
S_{\Psi}=\int \mathrm{d}t \mathrm{d}^3 x a^3\left(\sigma_1\dot{\Psi}^2-\frac{\sigma_2}{a^2}\vec{\nabla}^2\Psi\right).
\end{equation}
The coefficients $\sigma_1$ and $\sigma_2$ are defined as
\begin{eqnarray}
&&\sigma_1=3\beta_1+\frac{\beta_1^2}{\gamma_1^2}\gamma_2,
\\
&&\sigma_2=-\beta_2\left(1+\frac{H\beta_1}{\gamma_1}\right)
-\beta_1\frac{\mathrm{d}}{\mathrm{d}t}\left(\frac{\beta_1}{\gamma_1}\right).
\end{eqnarray}
Similar to the tensor modes, the coefficients $\sigma_1$ and $\sigma_2$ should be positive to avoid ghost and gradient instability.  Let us follow the method in section \ref{3}.
\bqn
\lb{scalar12}
\beta_1&=&1-2\hat{\alpha }\left(H-\frac{A}{a}\right)^2+4\hat{\alpha} \left(H-\frac{A}{a}\right)H,\nb\\
\beta_2&=&1+2\hat{\alpha }\left(H-\frac{A}{a}\right)^2-4\hat{\alpha}\left(-\dot{H}-\frac{A}{a}H\right),\nb\\
\gamma_1&=&-6\hat{\alpha}H^2\left(H-\frac{A}{a}\right)+6\hat{\alpha} H\left(H-\frac{A}{a}\right)^2\nb\\&&-2\hat{\alpha}\left(H-\frac{A}{a}\right)^3-H,\nb\\
\gamma_2&=&-3 H^2-4\dot{H}-2 \hat{\alpha}\left(\frac{A}{a}\right)^4-2\hat{\alpha} H^2 \left(3 H^2+4\dot{H} \right),\nb\\
\eqn
with $\Lambda_0=3H^2 +2\dot{H} +\hat{\alpha}\left(\frac{A}{a}\right)^4+3\hat{\alpha}H^4+4\hat{\alpha}H^2\dot{H}$.

\begin{figure}[htb]
\centering
\includegraphics[scale=0.4]{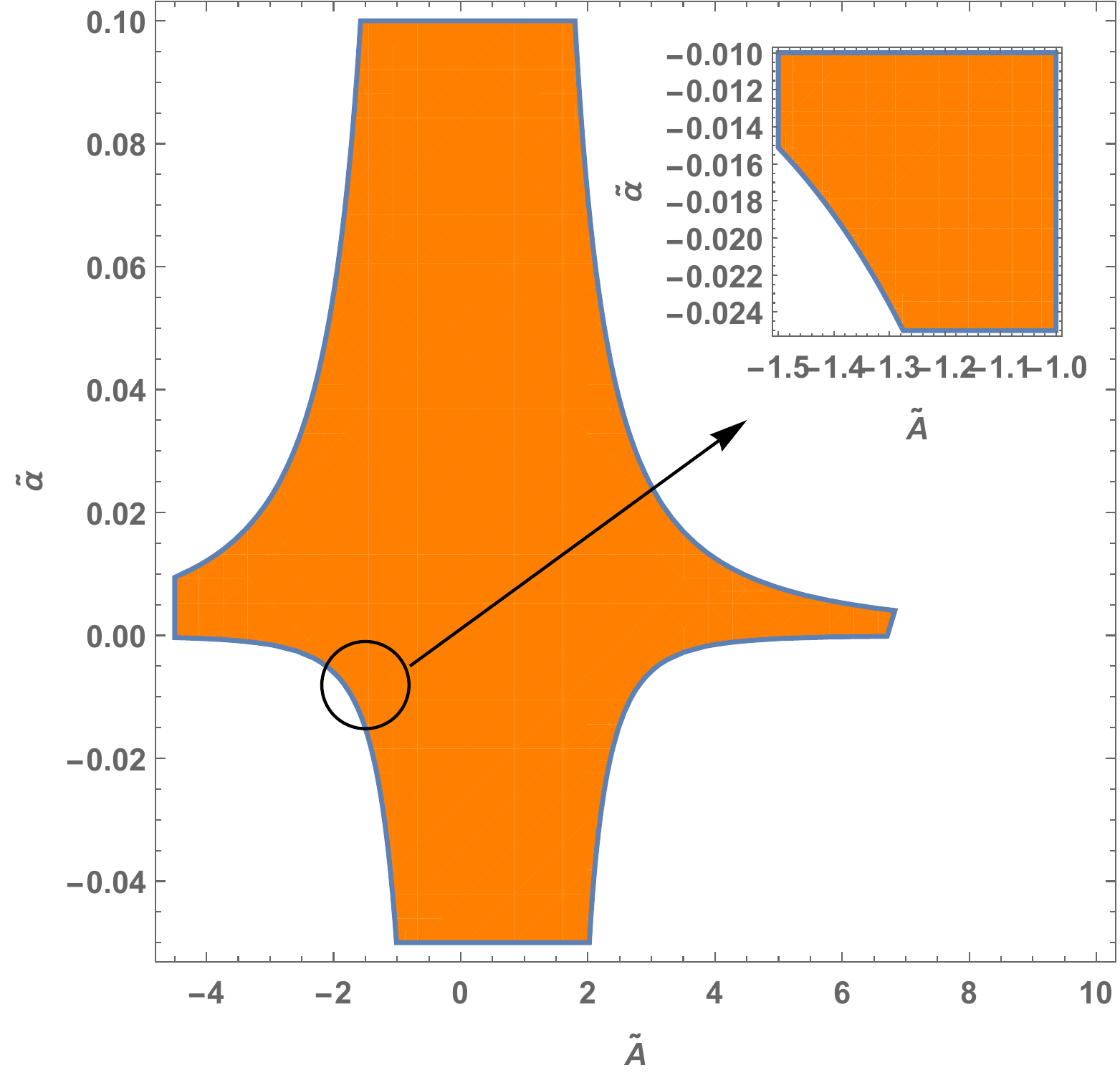}
\caption{Constraints on $\tilde{\alpha}$ and $\tilde{A}$ from theoretical  and observational bounds, the orange area satisfies $\sigma_1>0$ and $\sigma_2>0$.  The zoomed-in region is to show that around $-1.5<\tilde{A}<-1$, the current constraint obtained by $\sigma_1>0$ and $\sigma_2>0$ is much looser ($12$ order of magnitude looser) than the one obtained from GW observations as shown in Fig. \ref{fig2}.} \label{fig1}
\end{figure}

At present, substituting  \eqref{Hdot2} into \eqref{scalar12}, we have \eqref{beta12} and
\bqn
\gamma_1&=&\left(-6\tilde{\alpha}\left(1-\tilde{A}\right)+6\tilde{\alpha}\left(1-\tilde{A}\right)^2\right.\nb\\
&&\left.-2\tilde{\alpha}\left(1-\tilde{A}\right)^3-1\right)H_0,\nb\\
\gamma_2&=&\frac{6}{5}\left(5 \tilde{\alpha } \left(\tilde{A}^4-1\right)-1\right)H_0^2.
\eqn
Using $\sigma_1>0$ and $\sigma_2>0$, we get the constraints on $\tilde{\alpha}$ and $\tilde{A}_0$ as shown in Fig \ref{fig1}.

\begin{figure}[htb]
\centering
\includegraphics[scale=0.5]{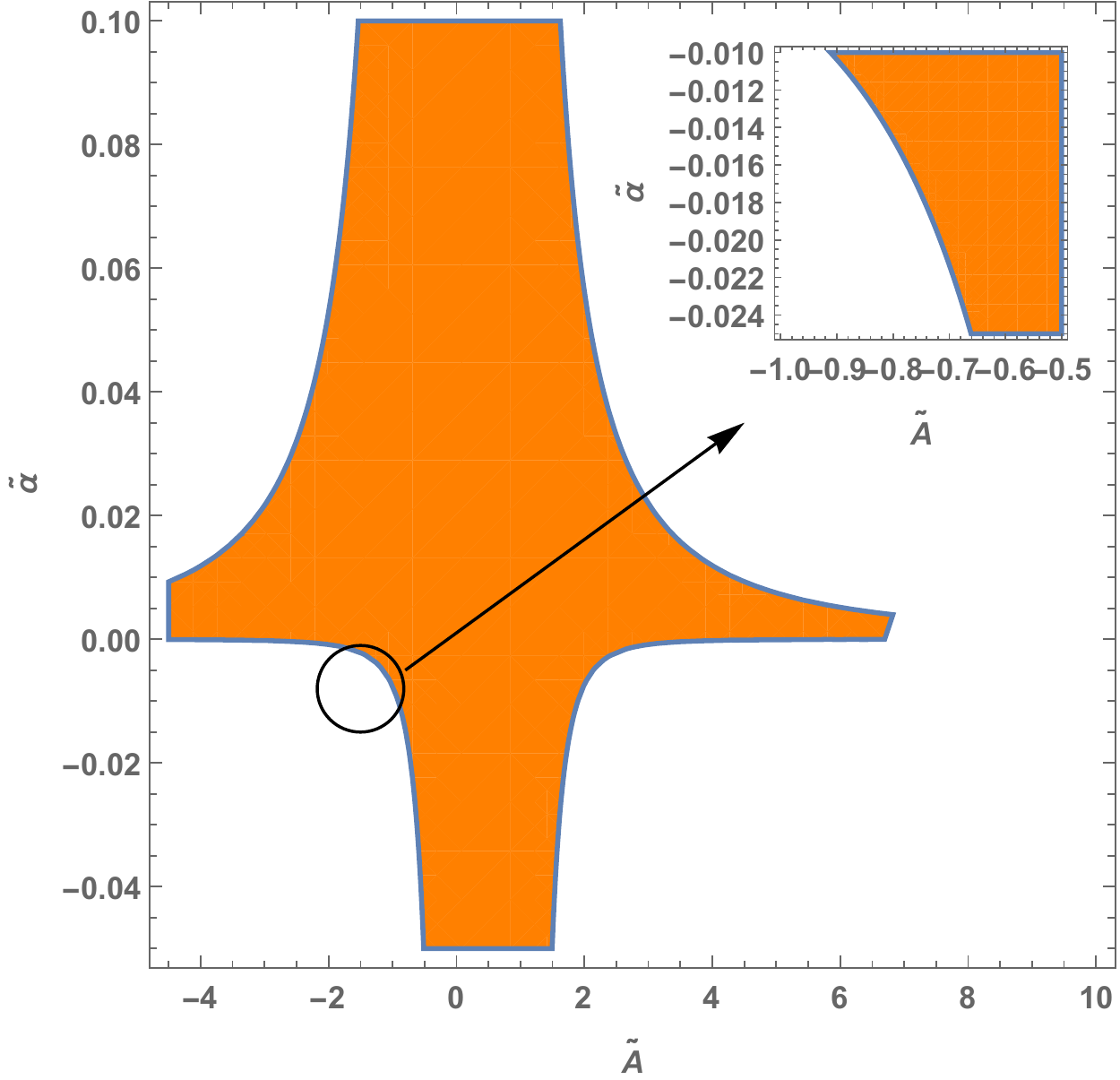}
\caption{For simplicity we set $\Omega_{c}=0.2589$. Then constraints on $\tilde{\alpha}$ and $\tilde{A}$ from theoretical  and observational bounds, the orange area satisfies $\sigma_1>0$ and $\sigma_2>\frac1{2}\left(\frac{\beta_1}{\gamma_1}\right)^2(\rho^{(\chi)}+p^{(\chi)})>0$. The zoomed-in region is to show that around $-1.0<\tilde{A}<-0.5$.} \label{fig1.1}
\end{figure}

In order to see whether the above results will be effected by coupling a matter field, in what follows we will consider the case where  matter field is mimicked by a k-essence field $\chi$ \cite{Kobayashi:2019hrl}. Following what we did for the case without $\chi$, we show that the quadratic action, after fixing the gauge, can be reduced to the modes $\Psi$ and $\delta \chi$ solely \cite{DeFelice:2011bh,Kobayashi:2019hrl}\bqn
S^{(2)}_M&=&\int \mathrm{d}t \mathrm{d}^3xa^3[G_{ij}\dot{Q}_i\dot{Q}_j- \frac1{a^2}F_{ij}(\partial Q_i)(\partial Q_j)\nb\\
&&-D_{ij}Q_i\dot{Q}_j-B_{ij}Q_iQ_j],\nb\\
\eqn
where
\begin{eqnarray}
  G_{ij} &=& \left(
    \begin{array}{cc}
      {\sigma_1}+Z & -Z \\
      -Z & Z
    \end{array}
  \right),
\quad
  F_{ij} = \left(
    \begin{array}{cc}
      {\sigma_2} & -c_m^2Z \\
      -c_m^2Z & c_m^2Z
    \end{array}
  \right),\nb\\
  Q_i&\equiv&\left(\Psi, \frac{\beta_1}{\gamma_1}\frac{\delta\chi}{\dot\chi} \right),
  \quad
  Z\equiv\left(\frac{\beta_1}{\gamma_1}\right)^2\frac{\rho^{(\chi)}+p^{(\chi)}}{2c_m^2}.
\end{eqnarray}Here $c_m^2\equiv \frac{\dot{\rho}^{(\chi)}}{\dot{p}^{(\chi)}}$ with $\rho^{(\chi)}+p^{(\chi)}>0$,  and $\rho^{(\chi)}$ and $p^{(\chi)}$ are respectively energy density, pressure of the fluid mimicked by k-essence field $\chi$. $D_{ij}, B_{ij}$ are the components of the 2 × 2 matrices \cite{DeFelice:2011bh}. Avoiding ghost instabilities requires that $\sigma_1>0$ and $Z>0$. Due to det$(v^2G_{ij}-F_{ij})=0$, the stability conditions including an additional perfect fluid become $\sigma_1>0$ and $\sigma_2>\frac1{2}\left(\frac{\beta_1}{\gamma_1}\right)^2(\rho^{(\chi)}+p^{(\chi)})>0$, in which $v$ is the propagation speeds of the two scalar modes. In summary, compared to the case without matter, the stability conditions are changed to $\sigma_1>0$ and $\sigma_2>\frac1{2}\left(\frac{\beta_1}{\gamma_1}\right)^2(\rho^{(\chi)}+p^{(\chi)})>0$, which depends on the values of $\rho^{(\chi)}$ and $p^{(\chi)}$.
As an example, suppose that $\chi$ represents dark matter. Let us use the Plank\textbf{ 2015 }data \cite{Aghanim:2018eyx}, where dark matter density parameter and radiation density parameter are $\Omega_{c}=0.2589±0.0057,$ $\Omega_R=0$. Then the constraints on  $\tilde{\alpha}$ and $\tilde{A}$ can be obtained as shown in Fig \ref{fig1.1}.
It is clear that the constraints on  $\tilde{\alpha}$ and $\tilde{A}$ are stricter by comparing with Fig \ref{fig1}. For the case that $\chi$ represents radiation, however, the enhancement of the constraints is negligible due to the ignorable radiation density. As a result, for this case we have almost the same constraints as those for the pure gravity case, i.e, the one without coupling matter field.

\section{Conclusions and Discussions}\label{sec:conclusion}

In this paper, we give detailed analysis about the theoretical and observational constraints on the regularized 4D EGB theory. Our analysis is based on linear perturbation around the FRW universe and is limited to the tensor modes such that we can deal with the gravitational waves. For these modes, the fluctuations of the scalar field $\phi$ are decoupled, and a set of linear perturbation equations are obtained, through which the speed of GWs can be read off.

Our results can be divided into two classes according to whether the bare cosmological constant $\Lambda_0$ is vanishing or not. For $\Lambda_0=0$, we find that theoretical requirements of the model are contradicted with the current observational results, indicating that the theory of this case should be ruled out. We make the conclusion from two strong evidences: one is from the theoretical contradiction with the current observations of $w_{de}$, the other comes from the huge (about 15 orders of magnitude) deviations between constraints from GW170817 and GRB 170817A and constraints from the current cosmological constraint on the ratio of energy densities between dark energy and matter, $\frac{\Omega_{de}}{\Omega_m}$. Including the integration constant $A$ in (14) does not remove the inconsistency obtained from the constraint of $w_{de}$ on $\tilde{\alpha}$ and $\tilde{A}$.

For $\Lambda_0\neq0$, however, one can place a stringent constraint on the dimensionless model parameter $\tilde{\alpha}$ (and $\Lambda_0$, since $\Lambda_0$ and $\tilde{\alpha}$ are not independent in this case as shown in \eqref{Lambda}). Compared to theoretical and cosmological constraints (values of $w_{de}$ and $\alpha_M$), the constraint from the speed of GWs measured by GW170817 and GRB 170817A is much more stringent. Specifically, it is given by $-7.78\times10^{-16}\le \tilde{\alpha} \leq 3.33\times10^{-15}$. In should be noted that in a modified version of the theory \cite{Aoki:2020iwm}, if the additional modification -- $k^4$ term of the dispersion relation is ignored, the model is effectively compatible with our model, and the speed of GWs place an upper bound $\alpha<10^{50}eV^{-2}$, which is equivalent to  $\tilde{\alpha}<10^{-16}$  in the our notation and is consistent with our result. In contrast, if the $k^4$ term is taken into considerations, the modified version of the theory \cite{Aoki:2020iwm} gives much stricter constraints.
As $A$ is taken into consideration,  we find that for $|\tilde{A}|\lesssim 1.35$ we have constraints $|\tilde{\alpha}|<3\times10^{-14}$. In addition, the case with $|\tilde{A}|\gtrsim 1.35$ is excluded by observational data as shown in Fig. \ref{fig2}.

Scalar perturbations in the universe dominated by $\phi$  has been considered in section \ref{4}. The Fig. \ref{fig1} shows that the current constraint obtained by $\sigma_1>0$ and $\sigma_2>0$ is much looser ($12$ order of magnitude looser) than the one obtained by GW observations as shown in Fig. \ref{fig2}. While including the other kind of matter,  the stability conditions are changed to $\sigma_1>0$ and $\sigma_2>\frac1{2}\left(\frac{\beta_1}{\gamma_1}\right)^2(\rho^{(\chi)}+p^{(\chi)})>0$, which depends on the values of $\rho^{(\chi)}$ and $p^{(\chi)}$. If $\chi$ represents dark matter, then the constraint on  $\tilde{\alpha}$ and $\tilde{A}$ will be stricter as shown in Fig \ref{fig1.1}.
\\

\textbf{Note added}: After this work was completed, we learned a similar work \cite{1803359}, which appeared in arXiv a few days before. Their work focused on $A=0$ case (where $A=0$ corresponds to $C=0$ in the notation of  \cite{1803359}). Besides, it is worth noting that much stronger constraints in \cite{1803359} have been obtained from other sources. For example, $-10^{10}m^2 \lesssim \alpha\lesssim 10^8 m^2$ from BH binary inspiral, $\alpha\simeq 10^{10} m^2$ from LAGEOS satellites and so on.

\section*{Acknowledgements}

Many thanks to Dong-Hui Du for helpful discussions. This work was supported in part by the National Natural Science Foundation of China under Grant Nos. 11975116, 11665016, 11947025 and Jiangxi Science Foundation for Distinguished Young Scientists under Grant No. 20192BCB23007.


\end{document}